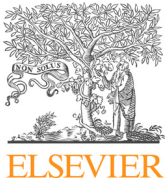

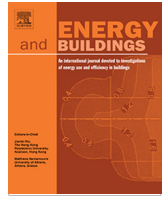

# A novel approach of day-ahead cooling load prediction and optimal control for ice-based thermal energy storage (TES) system in commercial buildings

Xuyuan Kang [a], Xiao Wang [a], Jingjing An [b], Da Yan [a,*]

[a] Building Energy Research Center, School of Architecture, Tsinghua University, Beijing 100084, China
[b] School of Environment and Energy Engineering, Beijing University of Civil Engineering and Architecture, Beijing 100044, China

ABSTRACT

Thermal energy storage (TES) is an effective method for load shifting and demand response in buildings. Optimal TES control and management are essential to improve the performance of the cooling system. Most existing TES systems operate on a fixed schedule, which cannot take full advantage of its load shifting capability, and requires extensive investigation and optimization. This study proposed a novel integrated load prediction and optimized control approach for ice-based TES in commercial buildings. A cooling load prediction model was developed and a mid-day modification mechanism was introduced into the prediction model to improve the accuracy. Based on the predictions, a rule-based control strategy was proposed according to the time-of-use tariff; the mid-day control adjustment mechanism was introduced in accordance with the mid-day prediction modifications. The proposed approach was applied in the ice-based TES system of a commercial complex in Beijing, and achieved a mean absolute error (MAE) of 389 kW and coefficient of variance of MAE of 12.5 %. The integrated prediction-based control strategy achieved an energy cost saving rate of 9.9 %. The proposed model was deployed in the realistic building automation system of the case building and significantly improved the efficiency and automation of the cooling system.

© 2022 Elsevier B.V. All rights reserved.

## 1. Introduction

Buildings are responsible for 29 % of the world's total energy consumption and 28 % of the world's total carbon emissions. [1] Global building energy consumption has increased by 48 % over the past 30 years, and carbon emissions have increased by 64 % [2]. Improving the energy efficiency and reducing carbon emissions has become a main goal throughout the world. Major countries have set ambitious targets to realize carbon neutrality by mid-21st century under the Paris Climate Agreement [3]. Introducing renewable sources such as photovoltaics (PV) [4], wind power [5], and hydropower [6], etc. in energy systems is one of the most efficient approaches to reduce carbon emissions in buildings. However, the instability and uncertainty [7] of renewable energy generation has brought great challenges for the demand–supply balancing in power grids. The "duck chart" of the California power grid is a typical example of the overgeneration from renewable energy sources and imbalance of supply and demands in the power grids [8]. Thus, demand response and demand-side management are essential solutions [9]. Implementing energy-flexible systems in buildings has been recognized as an effective approach to realizing demand response [10].

Thermal energy storage (TES) is a commonly used and effective system form to improve energy flexibility in commercial buildings. A typical ice-based TES system can charge the ice storage during off-peak hours at night and provide cooling during peak hours during the day [11]. The ice storage tank performs as a thermal battery to shift loads from the day to the night [12]. Moreover, TES has significant benefits on energy costs under the time-of-use (TOU) tariffs. A TOU tariff is a simple and effective approach to leverage peak-valley energy demands. It is a fixed pricing scheme that applies lower prices during the night and higher prices during the day, according to the balance of supply and demands in the grids [13]. The energy costs of chillers and cooling towers can be significantly reduced by managing the TES according to the TOU tariffs. Thus, TES, especially ice-based TES, has been widely applied to centralized cooling plants in large commercial complexes [14].

* Corresponding author at: Room 201, Building Energy Research Center, School of Architecture, Tsinghua University, Beijing 100084, China.
*E-mail address:* yanda@tsinghua.edu.cn (D. Yan).

Management and control are major factors in TES performance. TES capacity is typically designed to meet the total demand during peak TOU periods only (partial-peak TOU periods are not included) in design days. In most days, the ice storage cannot cover the total demand during the day. Therefore, it is critical to determine when to use TES or when to run chillers for cooling. As some peak TOU hours occur during later hours of the day, if the stored ice is used up early in the day and the chillers have to run during later peak TOU periods, the cost benefits will decrease. If the chillers run at some hours during the daytime and there is still remaining ice at the end of the day, the entire system has not taken full advantage of the TES system. Appropriate operating schedule is key in maximizing the energy cost saving potential with TES. The optimal TES management relies on two aspects: cooling load prediction and optimal control strategy.

The cooling load prediction model is the basis for TES management. Accurate cooling load prediction provides useful information for the future demand, thus supporting the decision making of cooling load allocations for different hours of the day. Cooling load prediction models have been studied by many researchers and have a series of well-acknowledged methodologies.

Traditional approaches of simulating and estimating cooling loads in buildings are based on building energy modeling programs (BEMPs). The research and development of thermal process-based BEMPs dates back to 1980 s and has produced a series of widely-used software including DeST [15], EnergyPlus [16], and ESP-r [17]. However, using BEMPs for cooling loads requires detailed inputs of building thermal properties and operating schedules, and usually needs calibration [18,19]. This approach may not be applicable for existing buildings owing to the large workloads [20] and possibly low predicting accuracy due to lack of precise inputs[21].

Most of the current studies adopt data-driven machine learning algorithms for step-ahead predictions [22]. Some research utilized neural-network-based algorithms for cooling load predictions. Fan et al. developed a series of short-term cooling load prediction models using ensemble learning [23], deep learning [24], and feature engineering [25], etc., and enhanced prediction model with transfer [26,27] and generative learning [28] techniques under various scenarios. Rahman et al. [29] optimized deep neural networks for medium- and long-term energy predictions in buildings. Some studies investigated the use of support vector machine algorithms. Zhong et al. [30] developed a vector field-based support vector machine model for office building hourly energy load predictions. Research has also been conducted by adopting decision tree-based ensemble methods [31]. Wang et al. [32] used and analyzed random forest model for hourly energy load predictions for educational buildings. Touzani et al. [33] investigated a gradient-boosting machine learning algorithm for commercial building energy predictions. Other studies compared and combined different categories of machine learning algorithms for load predictions. Wang et al. [34] compared the prediction performance of shallow machine learning and deep learning algorithms and concluded that Long Short Term Memory (LSTM) algorithm is recommended for short-term predictions and Extreme Gradient Boosting (XGB) for long-term predictions. Ahmad et al. [35] compared tree-based and neuron-based machine learning algorithms for high-resolution building energy load predictions and concluded an equally applicable performance in building energy analytics. Data-driven approaches have been extensively adopted [36] and optimized [37] for cooling, heating, or energy load predictions of all aspects in building performance analysis.

According to Hong et al. [38], there are approximately 10,000 research papers focusing on data-driven studies and applications in building energy analysis. Research has applied various types of machine learning algorithms to improve the accuracy of building energy load predictions [39]. However, most current studies only worked on the prediction models themselves, and seldom investigated how the model can be applied for different scenarios and how the models should be optimized in order to fit for different engineering purposes [21]. Especially for prediction-based control applications for TES systems, it is crucial to connect the prediction model with the control model, so that the prediction model can better support the control optimization according to the required control inputs, critical TES system features and advanced control mechanisms.

As for control optimizations, the control model is developed based on the cooling plant performance and predicted cooling load [40]. The output of the control model is the operating schedule of the chillers and TES. Substantial research has investigated the optimal control methods for building TES. Hajiah et al. [41] investigated optimal controls of using building thermal inertia and ice storage system to reduce energy costs in commercial buildings. Buttitta et al. [42] developed and compared a rule-based and MPC-based control strategy for building integrated thermal storage systems in residential building stocks that fitted for demand response initiatives for residential end-users. Kim et al.[43] also developed an MPC-based control optimization model and evaluate the performance in the demonstration project of an TES system to realize renewable integration and decarbonization. Luo et al. [11] optimized the management of an ice-based energy storage system with hourly cooling load predictions and Sequential Quadratic Programming optimizations. Henze [44] utilized a model-based predictive supervisory control for optimal control of building thermal mass and ice-based TES using TOU tariffs. Yu et al. [45] deployed an uncertainty-aware transactive control framework for TES optimal control under real-time energy prices. Sun et al. [46] reviewed the traditional and optimal strategies for peak load shifting control using TES system. Yu et al. [47] reviewed in depth the current methods for TES optimal control, including proportional-integral-derivative (PID) feedback control, model predictive control (MPC), neural network control, fuzzy control, reinforcement learning control, and hybrid control compared to traditional rule-based control.

Generally, current researchers have discovered some theoretical simulation-based approaches for optimized TES control using TOU tariffs for demand response [48]. However, the practical application that combines cooling load predictions and adopts in real TES cases is somehow limited. It is essential to develop an integrated model that combines cooling load predictions and strategy optimizations, and to be practically applied in real-time managements of the current TES operations, especially in commercial buildings.

Hence, most current research on load predictions and TES operational managements have not considered the complete approach that connects prediction and control strategy of TES in buildings. The integrated real-time improvement of combined prediction and optimization models have not been investigated. The applications of such integrated prediction and optimization models have not yet been evaluated in practical engineering cases.

This study proposed a novel approach of day-ahead cooling load prediction and optimal control strategy for ice-based TES in commercial buildings. The proposed approach is an integrated prediction and control optimization method with mid-day modifications and adjustments. The research is based on a commercial building and develops a day-ahead cooling load prediction model. Based on the predicted cooling loads, a rule-based control strategy is proposed according to the TOU tariff. Both prediction and control optimization models apply mid-day modification and adjustment technique correspondingly and improve the performance of the prediction and control. The proposed model is applied and validated in the case study, and the accuracy of the prediction model and the energy cost is compared across different scenarios. The model has been deployed in the real cases and further improvements of the research is discussed in detail.

## 2. Methodology

Fig. 1 illustrates the technical approach of this study. The methodology consisted of four parts: data preparation, cooling load prediction model, TES control optimization method, and model evaluations. The data preparation section describes the collection of cooling load and power consumption data from building automation system (BAS) along with the corresponding weather data from local weather station. The cooling load prediction model section establishes the general structure of the machine-learning-based prediction model and optimizes the model from perspectives including predicting algorithms, input feature selections, predicting mechanisms, and mid-day modifications. Based on the predicted cooling loads, the TES control optimization method then develops a rule-based control logic and adopts mid-day adjustments for control strategy in accordance with the mid-day modifications in cooling load predictions. The performance of the cooling load prediction model and TES control optimization method is evaluated with both mathematical metrics (mean absolute error, root mean square error, etc.) and engineering metric, specifically energy costs. The evaluation is conducted by comparing the performances of three scenarios and quantifies the improvement of model performances of the proposed model with baseline models.

The methodology details are explained in the following subsections.

### 2.1. Data preparation

The data used in this study was acquired from a commercial complex in Beijing, China. The case included a shopping mall (gross area 85,000 m²) and an office building (gross area 56,000 m²), both served by a same cooling plant. The cooling plant equips three duplex chillers (cooling capacity 4,395 kW), one normal chiller (cooling capacity 1,406 kW), and a centralized ice storage tank (ice storage capacity 76,000 kWh). The cooling load prediction and optimal control was based on the cooling system of this target building.

The case applied to TOU tariff of commercial buildings in Beijing, as shown in Fig. 2. The TOU profile included four periods: super-peak rates at 1.2145 RMB/kWh during periods of 11:00–13:00 and 16:00–17:00; peak rates at 1.0862 RMB/kWh during periods of 10:00–11:00, 13:00–15:00, and 18:00–21:00; partial-peak rates at 0.5675 RMB/kWh during periods of 7:00–10:00, 15:00–16:00, 17:00–18:00, and 21:00–23:00; and off-peak rates at 0.1001 RMB/kWh during period of 23:00–7:00. The rate during the day was around 10 times greater than that at night, and this feature benefits the TES to shift loads from days to nights and reduce significant energy costs. The management of the ice-based TES using this TOU tariff is the key issue to improve the general performance in this case.

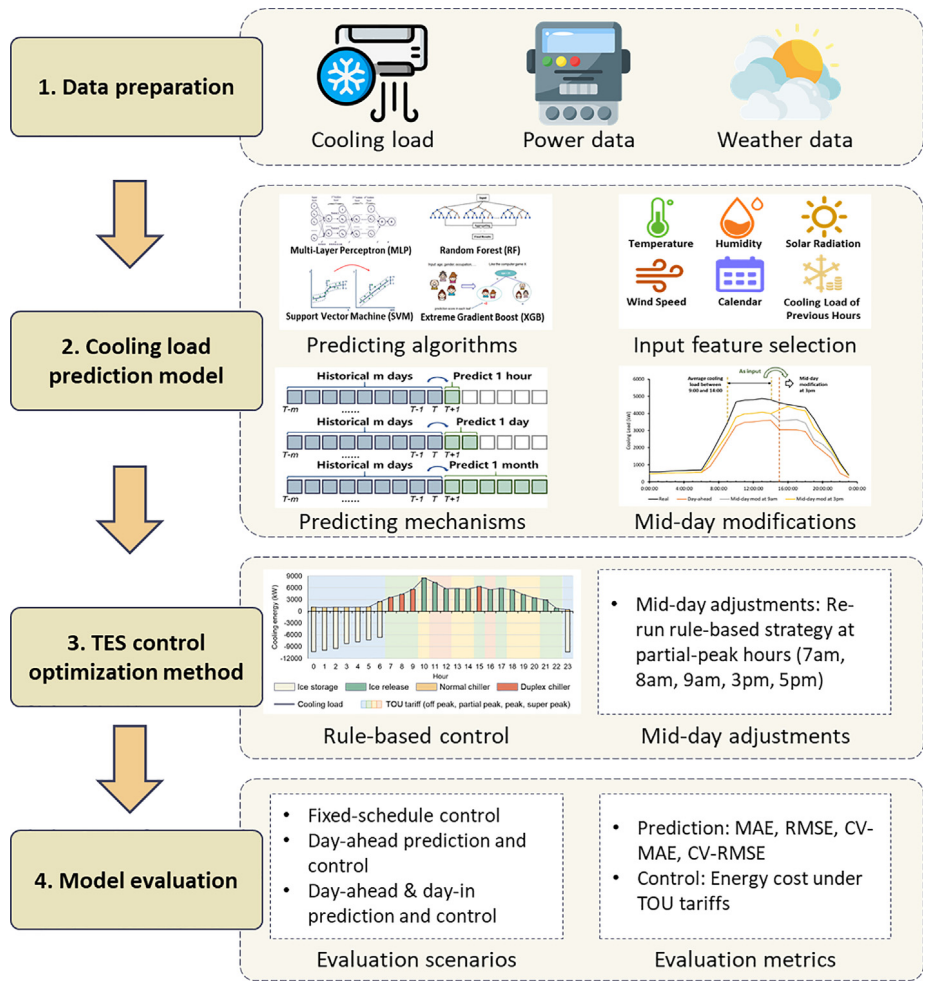

**Fig. 1.** Technical approach of the proposed research.

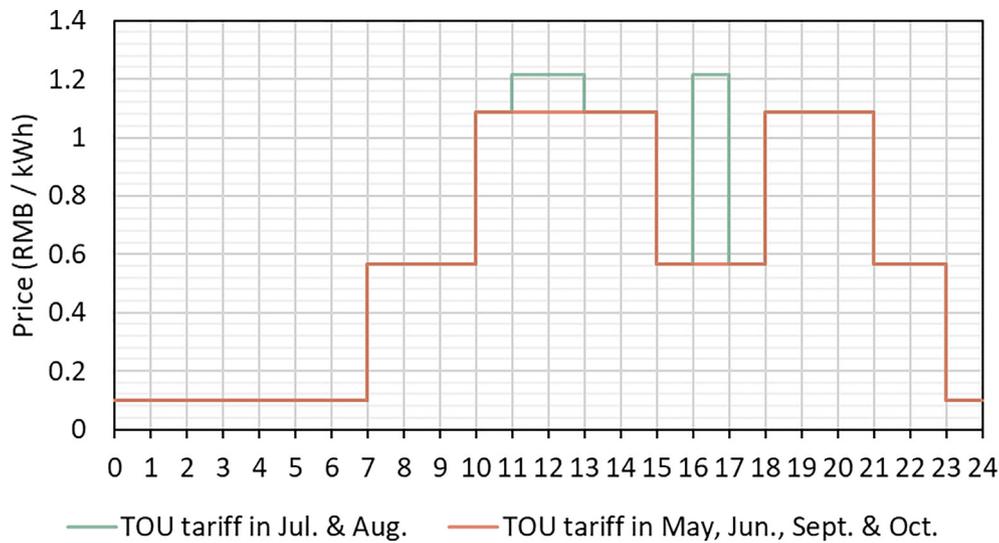

**Fig. 2.** Time-of-use (TOU) tariff for commercial buildings in Beijing, 2021.

The cooling load data for this study was calculated from hourly chilled water flow rate and supply/return water temperature, which were measured at the chilled water headers and acquired from the BAS. The sensors for these data have been calibrated and the validity of the data is cross-checked with balancing validation at different points of the cooling plant system. Together with the cooling load data, the daily electricity consumption data of chillers, pumps, and cooling towers were obtained from the BAS to support the cooling plant system model development. The cooling load data covered a period from July 1st to Oct. 21st, 2021. Weather data including ambient temperature, relative humidity, and wind speed, etc. of the corresponding period were collected. Historical weather data was acquired from ERA5 reanalysis dataset [49], while the real-time forecasted weather data was collected from an online API by ColorfulClouds service [50]. The calendar data marks days of the week for each day. The above-mentioned data was prepared and congregated for the development of cooling load prediction and optimal control models.

### 2.2. Cooling load prediction model

The cooling load prediction model uses machine learning methods as key algorithms, takes weather data, calendar information and historical cooling loads as inputs, and predicts cooling loads of future timesteps. The prediction model was trained and validated with the data of previous days, and predicts the cooling loads in a future window (a period of time). The prediction window then slides forward recursively and conducts the step-ahead predictions. The prediction model was developed and optimized from four perspectives: predicting algorithms, input feature selection, predicting mechanisms, and mid-day modifications.

### 2.2.1. Predicting algorithms

Several machine learning methods were selected and adopted as the core algorithms of the prediction model. These algorithms included multi-layer perceptron (MLP), support vector regression (SVR), random forest (RF), and extreme gradient boosting (XGB). The performances of the four algorithms were compared and the one with the lowest prediction error was used in the final prediction model.

MLP is a traditional machine learning algorithm that formulates a network of layers of non-linear approximators (neurons) and connects layers with linear combinations. The network is trained on a dataset using backpropagation and can be a powerful tool to build non-linear connections.

SVR is a powerful machine learning method that transform original samples into higher dimensions with kernel functions and constructs a hyper-plane for regression in high-dimensional space. SVR is a powerful tool that balances accuracy and the overfitting issue.

RF and XGB are both ensemble methods based on decision trees. RF builds and ensembles each decision tree with bootstrap sampling and bagging technique to aggregate the results of base predictors, while XGB ensembles decision trees with boosting techniques. Both methods are high-level complex algorithms for non-linear fitting and regression.

Training models with each of the above-mentioned methods requires hyperparameter tuning. The hyperparameters for each method are listed below in Table 1. Performance of the methods were calculated and compared to select the optimal algorithm for prediction model.

**Table 1**
Hyperparameters of the selected machine learning algorithms.

| Hyperparameters | | Code | Range of value |
|---|---|---|---|
| MLP | Number of layers | n_layer | [1, 2, 3] |
| | Number of neurons per layer | n_neuron | [10, 100, 500] |
| SVR | The type of kernel functions | kernel | ["poly", "rbf", "sigmoid"] |
| | Regularization parameter | C | [0.1, 1, 10, 100] |
| | Epsilon in the epsilon-tube | epsilon | [0.01, 0.1, 1, 10] |
| RF | Number of trees in the forest | n_estimators | [50, 80, 100, 150, 200] |
| | Maximum depth of trees | max_depth | [6, 8, 10, 12, 15] |
| | Minimum number of samples required to split an internal node | min_sample_split | [5, 10, 20, 49] |
| | Minimum number of samples required to be at a leaf node | min_samples_leaf | [2, 5, 10] |
| XGB | Maximum depth of trees | max_depth | [3, 4, 5, 6, 7] |
| | Minimum sum of instance weight needed in a child | min_child_weight | [0.01, 0.1, 1, 10] |
| | L1 regularization term on weights | reg_alpha | [0, 0.01, 1, 10] |
| | L2 regularization term on weights | reg_lambda | [0, 0.01, 1, 10] |
| | Objective function to be used | objective | ["reg: squarederror", "reg:tweedie"] |

### 2.2.2. Input feature selection

Input features in this case included ambient temperature (T), relative humidity (H), direct solar radiation (R), wind speed (S), weekday (W), hour (O), and historical cooling load (L). Weekday represents the day of the week of the samples. Mondays through Sundays are marked as 1 to 7 respectively, and holidays are marked as 8 according to the official holiday schedule in Beijing. Hour represents the hour of the day for each entry. Historical cooling load refers to the cooling load of the same hour in previous day, which marks the time relevance information of the cooling load profiles.

The final combination of the input features for the prediction model was selected from the above-mentioned seven features. The purpose of input feature selection is to find the optimal input feature combinations that contain features with positive effects on model accuracy. To serve this purpose, there are 127 ($2^7$-1 = 127) possible combinations of input features, and each combination was used to train and validate the prediction model. The input feature combination with the best prediction performance was used in final model. This approach is an exhaustive search of optimal input feature combination and selects the best input features for prediction.

### 2.2.3. Predicting mechanisms

Predicting mechanisms include two aspects: predicting window and historical data size. As this research applies step-ahead sliding window prediction, the prediction window refers to the duration of the forecasted cooling load in the following step before re-train the model (as illustrated in Fig. 3(a)). The historical data size refers to the size of the data required for training and tuning the prediction model, specified as the size of consecutive historical several days (as illustrated in Fig. 3(b)). Predicting window and historical data size both need tuning. The prediction window was selected among 1 day, 1 week and 1 month, while the historical data size was selected among 7, 14, 30 and 60 days. Models under each mechanism are trained and the optimal configuration was selected regarding predicting accuracy.

### 2.2.4. Mid-day modification

Traditional cooling load prediction follows day-ahead predicting mechanisms. This is in line with the need of day-ahead scheduling of chiller operations and ice-based TES management. However, the uncertainty of the cooling load may result in significant variance of the predictions, thus making the control strategy less effective or efficient. As a result, we introduced a mid-day modification mechanism to modify the predicted cooling loads at certain hours in the middle of the day based on the observed cooling loads of the previous hours. The factural cooling load data in the near past provides supportive information for the cooling load consumptions.

The mid-day modification was conducted for five times at 7 am, 8 am, 9 am, 3 pm, and 5 pm, respectively. An illustration of mid-day modification mechanism at 9 am and 3 pm is shown in Fig. 4. The modification timepoint was selected according to the TOU tariff characteristics in Beijing. Since the stored amount of ice can normally meet the cooling demands during super-peak periods and peak periods on most days, the trade-off of control strategy usually occurs during partial-peak periods. The remaining amount of ice can only meet the cooling demand during some of the partial-peak hours. The trade-off strategy during partial-peak periods depends on the predicted cooling load of this period. As a result, accurate cooling load prediction during partial-peak periods is of immense significance, and the mid-day modifications are allocated right before the major partial-peak periods in the middle of the day to improve the overall performance of the cooling load prediction models and subsequent control strategy optimization model. Thus, the mid-day modification is conducted at every partial-peak hour except 9 pm and 10 pm. Since 9 pm and 10 pm are latest partial-peak hours in the day, it is definitely beneficial to use ice storage if there's remaining ice. These two hours require no critical trade-offs and mid-day modification is not necessary in these two hours. The modification hours are selected according to the above-mentioned principles.

In the prediction model for mid-day modification, the average cooling load of the hours between current and previous the modification timepoints was also introduced as an additional input. For example, the mid-day modification at 7 am takes the average cooling load from 12 am to 6 am as an additional input, while the mid-day modification at 3 pm takes the average cooling load from 9:00 to 14:00 as an additional input. A new mid-day prediction model was trained with such input data and predicted the cooling loads during hours of the rest of the day. The modified cooling load overrode the previous predictions and was used for subsequent mid-day control logic adjustments.

## 2.3. Control optimization method

The predicted cooling load was used for optimal control of the ice-based TES in commercial buildings. The key to the optimization problem is to determine when to use ice storage system during different hours of the day. The capacity of the ice-based TES cannot meet the total daily cooling demands on most days. According to the TOU tariff in Beijing, some of the peak or super-peak hours occurs during later periods of the day (16:00–20:00). If a more conservative strategy applies, the chillers will be used to cover the cooling demand for more hours in the day and may leave unused ice at the end of the day. If a more aggressive strategy applies, more ice may be used during early partial-peak hours and may result in running chillers during peak or super-peak periods in the later hours. Both scenarios may reduce the benefit of the TES system. The control optimization strategy aims to determine the chiller and TES operation schedule in the day based on predicted cooling load, in order to achieve greater energy cost savings.

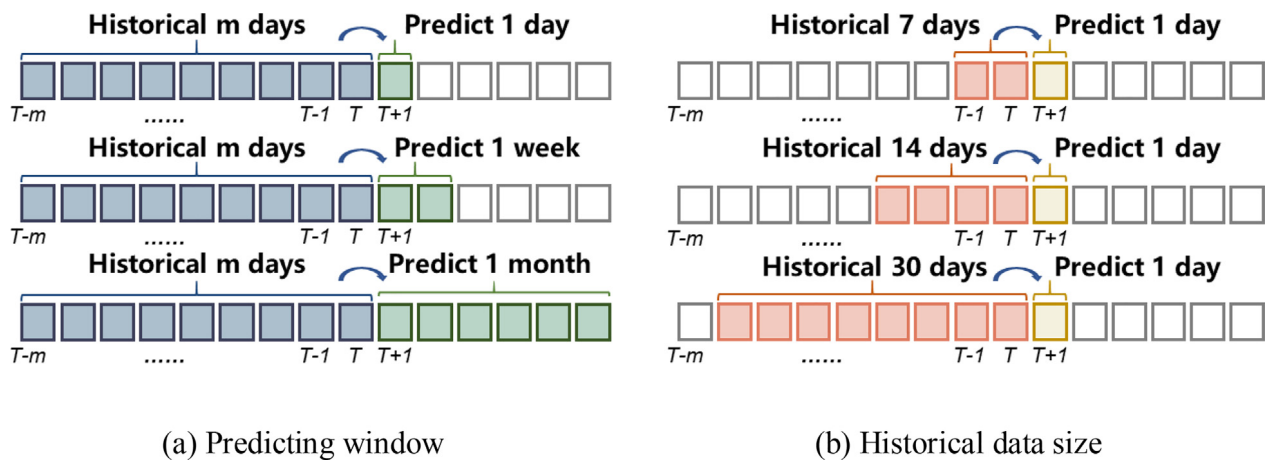

(a) Predicting window            (b) Historical data size

**Fig. 3.** Illustration of predicting window and historical data size.

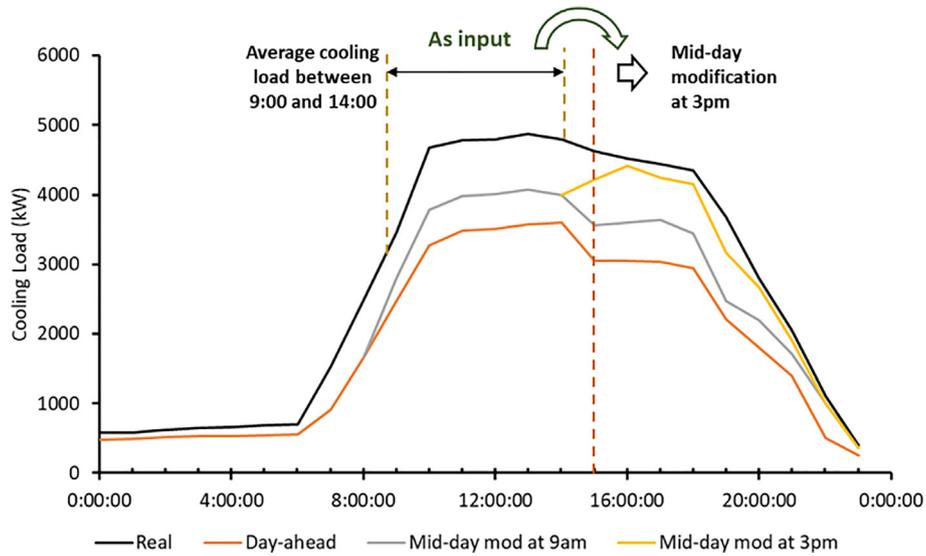

**Fig. 4.** Illustration of mid-day modification for cooling load predictions.

The optimal control logic is a rule-based strategy, as shown in Fig. 5. The decision-making process is based on the stored amount of ice ($s\_ice$) and predicted hourly cooling load ($cl$) in the day, which shall be inputted into the control logic. An hour sequence ($hour\_seq$) is determined and the stored amount of ice is allocated by the hour sequence according to the predicted cooling load. Following the rule-based control logic, a control strategy series ($ctrl$) is determined with 0 representing using stored ice and 1 representing using chillers for cooling demands in the building.

Within this rule-based control logic, the hour sequence is crucial as it contains the rank of priority for each hour. Taking the TOU tariff in July and August for commercial buildings in Beijing, the super-peak hours have the top priority that the stored ice should be first allocated to meet the demands in these hours (Hour 11, 12, 16). The remaining ice should then be used to meet the demands in the peak hours (Hour 10, 13, 14, 18, 19, 20). The remaining ice after the two-round allocation will be assigned to partial-peak hours. There are generally-three partial-peak periods: early mornings (Hour 7, 8, 9), late afternoon (Hour 15, 17), and late evening (Hour 21, 22). The remaining ice will be assigned firstly to late afternoon period (Hour 17, 15), then early morning period (Hour 9, 8, 7) in a reversed chronological order. If the stored ice is allocated chronologically, and the ice is firstly used at Hour 7, 8, 9, there are risks that there will not be sufficient ice during later peak hours if the predicted load is generally lower than the real load. Therefore, the reason of following a reversed chronological order is to reserve as much ice as possible for later hours and to avoid forced chiller operations during peak hours due to the variance of cooling load prediction. The late evening period is of the least priority because in real-time management, if there still exists remaining ice at the start of Hour 21, it is always beneficial to use stored ice to meet the cooling demand at the end of the day. The

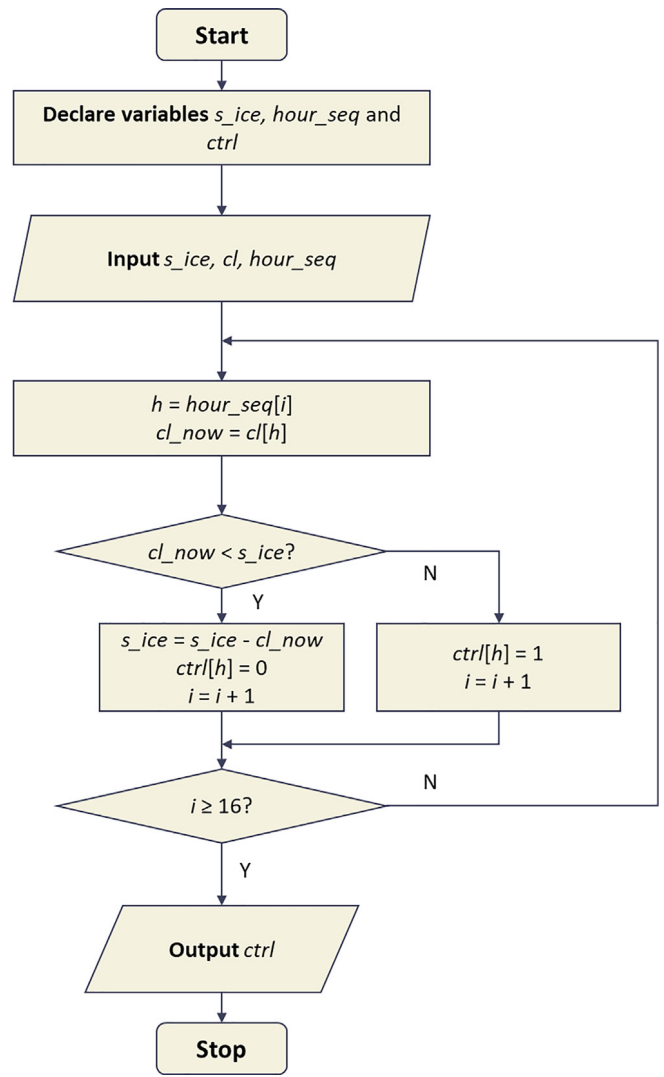

**Fig. 5.** Rule-based control strategy for optimal control of TES.

final hour sequence is then determined as: Hour 11, 12, 16, 10, 13, 14, 18, 19, 20, 17, 15, 9, 8, 7, 22, 21. The hour sequence under other TOU tariff profiles can be similarly determined using the above principles. This measure tends to conduct a more conservative approach to allocate stored ice in order to meet the demands and reduce energy costs during the day.

In addition to the general rule-based control strategy, a mid-day control adjustment strategy was also introduced in accordance with the mid-day prediction modifications. The mid-day prediction was conducted at Hour 7, 8, 9, 15, 17. According to the updated predictions, the control strategy may be adjusted following the similar rule-based control logic. The only variation is the hour sequence (*hour_seq*). Previous hours before the timepoint of adjustment was removed from the hour sequence, and the stored amount of ice was renewed according to the remaining ice at the timepoint of adjustment. The mid-day control adjustments used updated predictions as future indicators, and made decisions based on real-time remaining amount of ice, thus providing more efficient control strategy in the next hours. The mid-day modification and adjustment technique is an essential part to integrate prediction and control optimization model for advanced TES operation managements.

### 2.4. Model evaluation

The cooling load prediction model was evaluated and validated following the commonly used metrics and mechanisms. The cooling load predictions were evaluated with mean absolute error (MAE), coefficient of variance of MAE (CVMAE), root mean squared error (RMSE), and coefficient of variance of RMSE (CVRMSE). The definitions of these metrics are shown below:

$$MAE = \frac{1}{n} \sum_{i=1}^{n} |X_i - \widehat{X}_i| \qquad (1)$$

$$CVMAE = MAE / \bar{X}_i \qquad (2)$$

$$RMSE = \sqrt{\frac{1}{n} \sum_{i=1}^{n} \left( X_i - \widehat{X}_i \right)^2} \qquad (3)$$

$$CVRMSE = RMSE / \bar{X}_i \qquad (4)$$

The validation of cooling load prediction model follows the k-fold cross validation. The original training set is divided into k equal-sized subsets. In each iteration, one of the subsets is retained for validation and the other k-1 subsets are used to train the prediction model. The averaged prediction error for each validation subset is used as the final prediction error of the model. In this research, a 5-fold cross validation mechanism was applied.

The evaluation metric for optimal rule-based control method is the total energy cost during the cooling season. The energy cost is calculated using the commercial building TOU tariff in Beijing (see Fig. 2). As a result, simulating the hourly electricity consumption based on the real cooling demands and different control strategies is essential for energy cost calculation.

To simulate the hourly electricity consumption, a cooling plant system model was developed for the building. Key components of the system include: 3 duplex chillers, 1 normal chiller, 1 cooling tower group containing 4 cooling towers, 1 ice storage tank and multiple pumps. The simulation model for each component was established and calibrated. The chiller model was established by regressing chiller COP with cooling water temperature, chilled water temperature, and partial load rate using quadratic functions. The regression was based on the part-load data of the chiller sample curve. The cooling towers use constant-frequency fans and thus operate at rated power when the cooling tower is on. The ice tank is simulated by modeling the thermal transfer and the process of liquefaction or solidification of the water in ice balls. The detailed thermal modeling of the ice tank affects the performance of the duplex chillers and ethylene glycol pumps. The performance of the pumps is modeled based on the pump characteristic curves from the pump samples. The electricity consumption of the pumps is determined by the flow rates of the chilled water, cooling water or ethylene glycol respectively. The flow rates and balances of the chilled water loop, cooling water loop and ethylene glycol loop are also modeled to support the detailed simulation of chillers and cooling towers.

The parameters of models for different components in the cooling plant system are calibrated and validated using the measured electricity consumption data collected from the BAS. The total electricity consumption is a combination of the electricity consumption of chillers, cooling towers, and pumps. Through validation, the mean absolute percentage error of the total electricity consumption of the cooling plant system was 5.77 %. The accuracy of the cooling plant system model was sufficient for the modeling and calculation of the energy cost in this study.

The cooling load prediction and control optimization model performances were evaluated by comparing three scenarios including:

(1) Fixed-schedule scenario. The operation of the ice-based TES and chillers follows a fixed schedule; In July and August, the duplex chillers run at hour 7, 8, 9, 15, 17 and 22, and the ice-based TES supplies chilled water for the rest of the hours during the day. While in May, June, September and October, the duplex chillers only run at hour 7 and 22, and the cooling of the other hours during the day is supplied by ice-based TES. This schedule is the baseline scenario and also the currently applied schedule in the building;

(2) Day-ahead prediction and control scenario. In this scenario, the cooling load prediction results are from the day-ahead prediction model, without the mid-day modifications. The control strategies are also determined by running the control optimization model ahead of the day, without mid-day adjustments. The day-ahead-only scenario is widely applied in most ice-based TES system cases;

(3) Mid-day prediction and control scenario. In this scenario, both day-ahead predictions and mid-day modifications for cooling loads are applied. The control strategy is also determined and adjusted both ahead of the day and in the middle of the day correspondingly. The mid-day prediction modifications and control strategy adjustments conduct at hour 7, 8, 9, 15 and 17. This scenario uses the proposed method in this study.

The error of the cooling load prediction models and the energy costs of the control optimization methods under the above-mentioned three scenarios were compared and evaluated. The results of the evaluation and comparison are explained in Section 3.

## 3. Case study

### 3.1. Cooling load prediction model development and optimization

The approach of the establishment and optimization of cooling load prediction model was applied to the above-mentioned commercial complex case. To improve the performance of cooling load prediction model, the predicting algorithms, input features, histor-ical data size, and prediction window were carefully selected and optimized. Moreover, the modification of mid-day predictions was implemented to improve the accuracy of cooling load prediction. The cooling load prediction models for the shopping mall and the office building in the commercial complex were separately established and trained. The prediction performance of the optimized model was evaluated with the cooling load data of the commercial complex. It is to be noted that, the following section 3.1.1, section 3.1.2 and section 3.1.3 were all the results from simple day-ahead prediction models. Evaluations were based on predicted daily cooling load profiles. The section 3.1.4 introduced a mid-day modification mechanism, and the evaluations were based on the results of combined cooling load profile with the latest modified predictions.

### 3.1.1. Predicting algorithm selection

Four algorithms were compared: SVM, MLP, XGB, and RF. In the comparison of different algorithms, the input features were set to include ambient temperature, relative humidity, direct solar radiation, wind speed, hour and historical cooling load (THRSOL) for shopping mall and ambient temperature, relative humidity, direct solar radiation, weekday, hour and historical cooling load (THRWOL) for office building. The historical data size was set to 60 days and predicting window was set to 1 day. The settings of the remaining steps were all based on this initial set and were updated with the optimal selections. The prediction errors of the algorithms in MAE are shown in Fig. 6. The algorithm of RF was selected as it has the lowest MAE among the four algorithms in both the shopping mall (343 kW) and office building (203 kW).

The data experiment was conducted on a computing server with 20 cores (Intel Xeon® @ 2.3 GHz) and 64 GB RAM. The average computation time of each prediction in training process for different algorithms was also recorded. For the shopping mall, it costs 23.30 s on average to train the model and predict next 24 h cooling load, while for the office building, the process costs 22.68 s (Fig. 7). The computation time for predicting is less than 1 s, which is sufficient for real-time prediction computations. The computation time of RF was moderate among the four algorithms, and it is acceptable in practical engineering application. Therefore, RF was selected as the prediction algorithm of the cooling load prediction model.

### 3.1.2. Input feature selection

Seven input features were considered, including ambient temperature (T), relative humidity (H), direct solar radiation (R), wind speed (S), weekday (W), hour (O), and historical cooling load (L).

All combinations of the 7 input features ($2^7 - 1 = 127$ combinations in total) were tested and the combination with the lowest MAE was selected. Fig. 8 demonstrates the results of 7 combinations for shopping mall and office building to illustrate the contribution of each input feature intuitively. The full results of all 127 combination cases are listed in Table A1 (see Appendix A). For the shopping mall, the input feature combination of ambient temperature, relative humidity, direct solar radiation, wind speed, hour, and historical cooling load (THRSOL) had the lowest MAE of 343 kW among all 127 combinations. The input feature of weekday (W) lead to an increase of MAE. The most accurate input feature combination of office building was ambient temperature, relative humidity, direct solar radiation, weekday, hour, and historical cooling load (THRWOL), whose MAE was 203 kW.

### 3.1.3. Predicting mechanism setting

The predicting mechanism has been illustrated in two aspects: historical data size and predicting window. For the historical data size, Fig. 9 shows the comparison of four different historical data sizes including 7, 14, 30, and 60 days. The results suggest that training the cooling load prediction model with 60-day hourly historical data reaches the lowest MAE among the four historical data sizes for both the shopping mall (343 kW) and office building (203 kW). The historical data size was set to 60 days with respect to the prediction accuracy.

The prediction window sizes of 1, 7, and 30 d are compared as shown in Fig. 10. 1 day ahead prediction performs the lowest prediction error for the shopping mall (343 kW) and office building (203 kW), thus the prediction window size is set to 1 day.

### 3.1.4. Mid-day modification

With the above steps of prediction algorithm selection, input feature selection, and predicting mechanism setting, a classic day-head cooling load prediction model was established to predict the future 24 h cooling load at 12 am on each day. Furthermore, the Mid-day6 model was proposed to compensate the prediction-based TES cooling system control.

In the Mid-day6 model, 5 other predictions in the day (at 7 am, 8 am, 9 am, 3 pm, and 5 pm) were made in addition to the day-ahead prediction at 12 am. A Mid-day24 model was also developed to represent the most delicate case for mid-day modification. The historical cooling load of the last hour (equivalent to the average hourly historical cooling load from the previous prediction timepoint to the current timepoint) was used as an additional input feature. Predictions were made at each hour of the day to obtain the predicted cooling load of the remaining hours of the day.

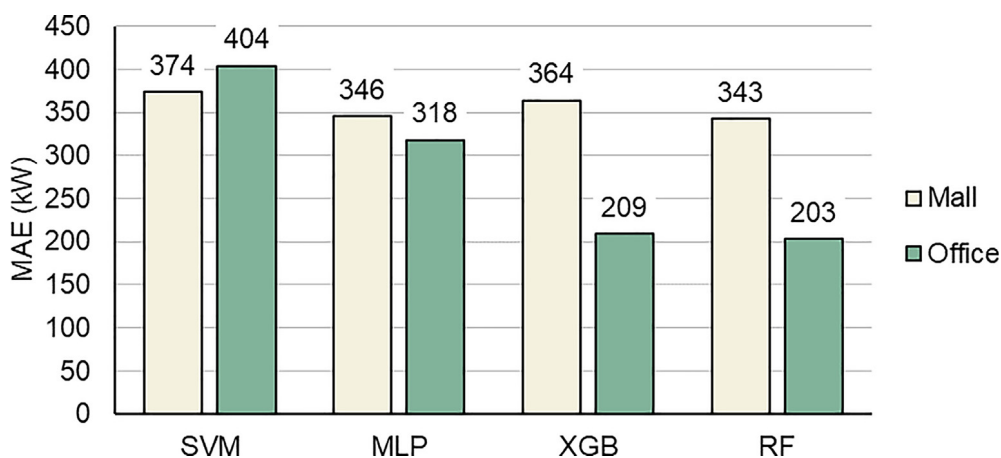

**Fig. 6.** Comparison results of different algorithms for Mall and Office.

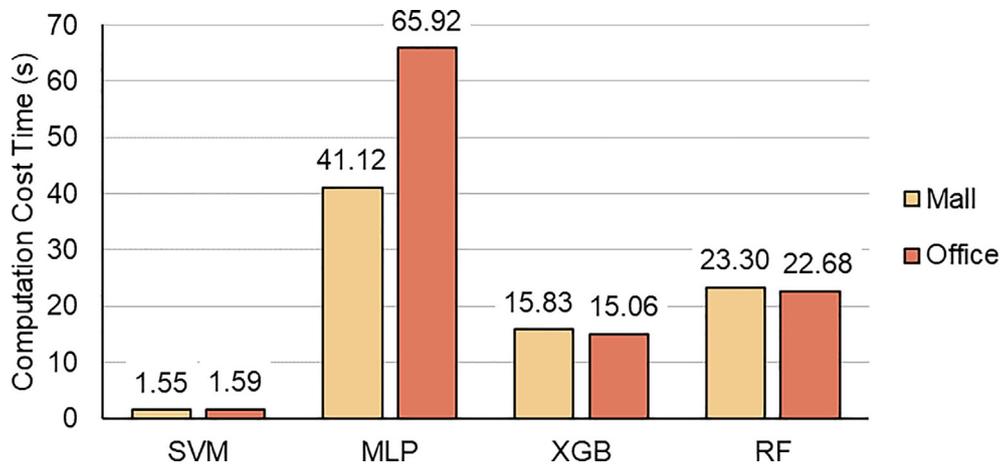

**Fig. 7.** Computation time in training process of different predicting algorithms for Mall and Office.

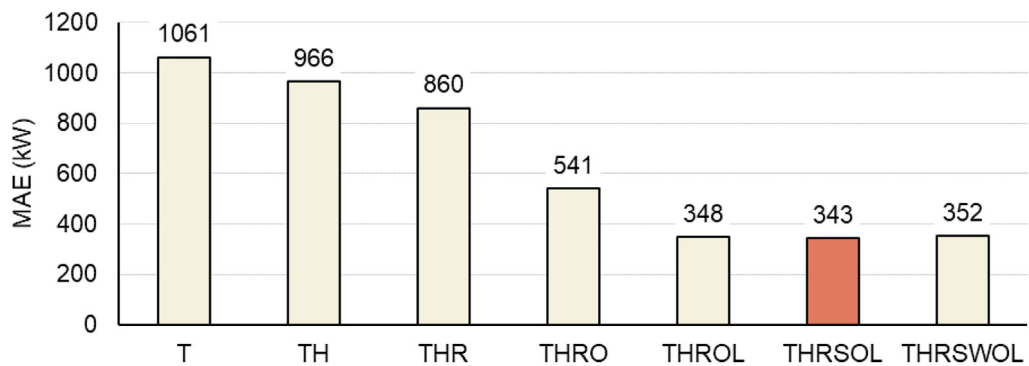

*(a) Mall*

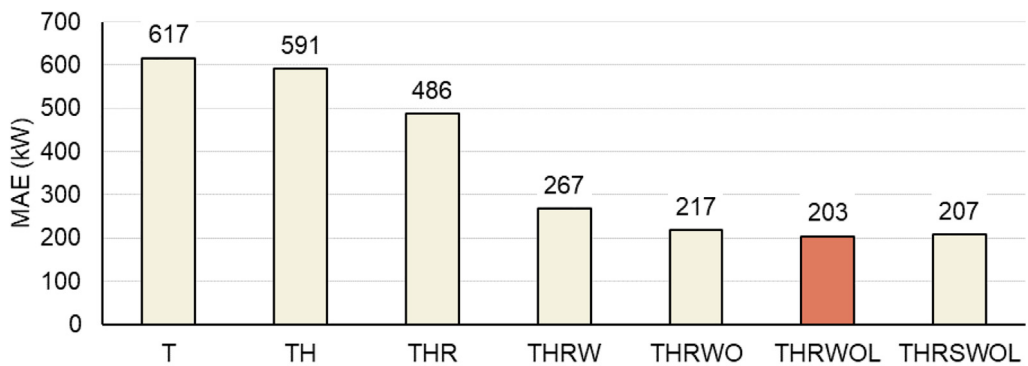

*(b) Office*

**Fig. 8.** Comparison of prediction error with different input features for Mall (a) and Office (b).

Fig. 11 shows the prediction performance of total cooling load of the shopping mall and office building on a selected day (July 21st, 2021), when the cooling load profile changed dramatically in the morning hours. With the mid-day modifications, the prediction result of the Mid-day6 model was more accurate than that of the Day-ahead model. On July 21st, 2021, the MAE of the Day-ahead, Mid-day6, and Mid-day24 models were 1,717 kW,

1,157 kW, and 518 kW, respectively. The Mid-day24 model updates hourly and can track the change of the cooling load profile timely, hence having a significantly better performance than the Mid-day6 model on the selected day.

The cooling load prediction result of the entire cooling season was compared among the proposed Mid-day6 model, Mid-day24 model with the simple model, and day-ahead model. Specifically, the simple model used the 24 h real cooling load of the previous day as the predicted 24 h cooling load. Table 2 demonstrates the overall prediction error of the whole dataset and the CVMAE of the Mid-day6 model is 12.5 %. Fig. 12 illustrates the hourly prediction results of the week from July 19th, 2021 to July 26th, 2021.

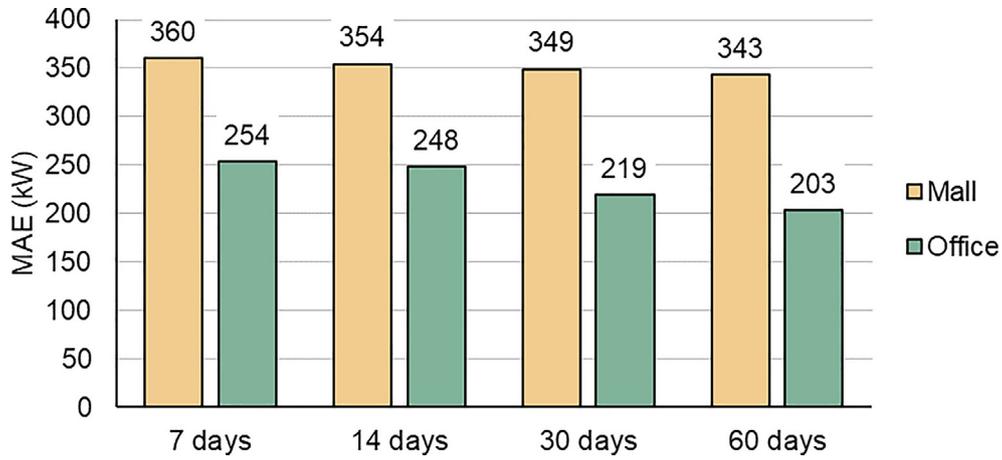

**Fig. 9.** Comparison results of historical data size for model training.

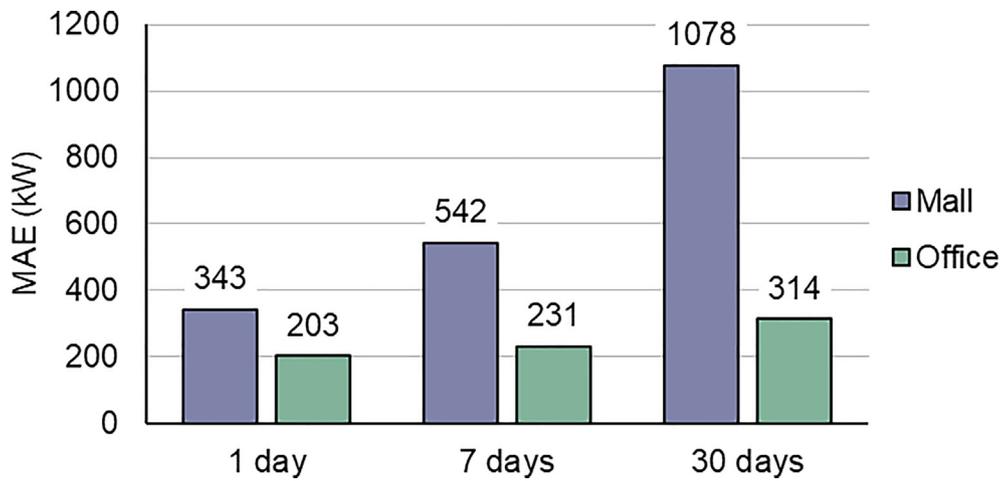

**Fig. 10.** Comparison results of prediction window size for Mall and Office.

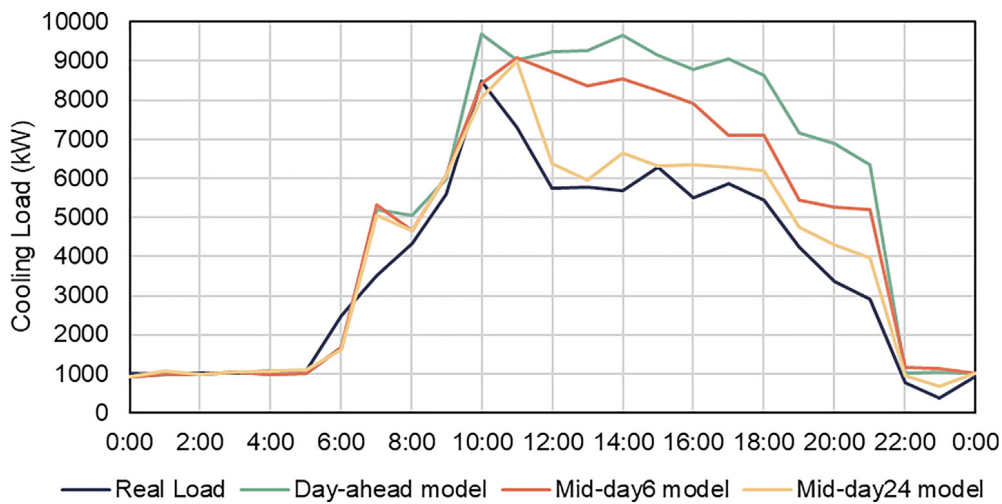

**Fig. 11.** Mid-day modification result on July 21, 2021.

The mid-day modification operation had an obvious prediction improvement compared to that of the day-ahead model. The proposed Mid-day6 model tracks the trend of cooling load of the commercial complex and predicts the cooling load accurately. The difference between the Mid-day6 and Mid-day24 models was minor as the Mid-day6 operations were sufficient to track some of the key points of the cooling load profiles in the day. This difference on prediction results bring about a slighter effect on the control results, which are analyzed in depth in the following subsection.

**Table 2**
Overall prediction errors of the four models.

|  | Simple model | Day-ahead model | Mid-day6 model (Proposed model) | Mid-day24 model |
|---|---|---|---|---|
| MAE (kW) | 577 | 452 | 389 | 340 |
| CVMAE | 18.5 % | 14.5 % | 12.5 % | 10.9 % |
| RMSE (kW) | 944 | 712 | 599 | 558 |
| CVRMSE | 30.3 % | 22.9 % | 19.2 % | 17.9 % |

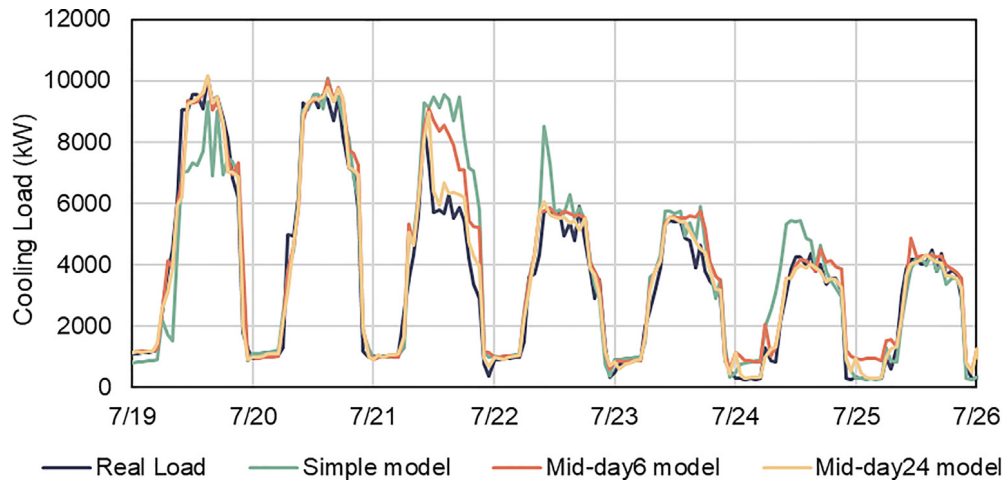

**Fig. 12.** Prediction results of the four models from July 19th to July 26th, 2021.

### 3.2. Control optimization and data-experimental results

Based on the cooling load prediction, the TES system hourly control strategy was determined using the rule-based control method. The energy simulation of the TES system was conducted with the previously established TES physical model of this commercial complex. The energy cost of the TES system was analyzed under three different control methods: Fixed schedule, Day-ahead prediction-based control, and Mid-day (Mid-day6) prediction-based control. The energy cost under the three control methods were calculated according to the TOU tariff of Beijing.

Two specific days were selected to demonstrate the control performance of the proposed prediction-based control method for the TES system. One was July 21st, 2021, when the cooling load is generally high and there is a super-peak in TOU tariff. The other was Sept. 7th, 2021, when the cooling load was relatively low and the TOU tariff operated without super-peak periods.

Fig. 13 shows the control performance on July 21st, 2021. At off-peak hours, the three strategies operate duplex chiller for ice storage and use normal chiller to cover the cooling demand of the commercial complex. At peak and super-peak hours, the three strategies release ice for cooling supply. At partial-peak hours, the difference of the three control strategies is whether to use duplex chiller or stored ice to cover the cooling load. The Fixed schedule used duplex chiller for cooling supply at each partial-peak hour except 9 pm, which was relatively conservative. With cooling load prediction, Day-ahead prediction-based control used ice at 10 pm instead of duplex chiller. The electricity cost under Day-ahead prediction-based control (10,936 RMB) was lower than that under Fixed schedule (11,212 RMB) on July 21st, 2021. The modification of Mid-day cooling load prediction improves the prediction accuracy (Fig. 11), which supports to make a more cost-effective decision. Compared to Day-ahead prediction-based control, Mid-day6 prediction-based control used ice at 5 pm. The overall electricity cost was reduced to 10,055 RMB on July 21st, 2021. The Mid-day control optimization model generally performed more cost-effectively on hot summer days.

The control performance on Sept. 7th, 2021 is illustrated in Fig. 14. The Fixed schedule used duplex chiller at 7 am and 10 pm, while Day-ahead prediction-based control and Mid-day6 prediction-based control used ice at these hours. The operation strategy of the three control methods was the same at other hours. The electricity cost of Day-ahead prediction-based and Mid-day6 prediction-based controls had a minor difference owing to ice storage during off-peak hours, which is caused by the ice usage of previous days. The electricity cost on Sept. 7th, 2021 was 6,100 RMB, 5,700 RMB, and 5,590 RMB for Fixed schedule, Day-ahead prediction-based control, and Mid-day6 prediction-based control, respectively. Generally, the Mid-day optimization method operated similarly with the Day-ahead optimization method and had moderately better performance than the Fixed-schedule control on days with mild weather.

The total energy cost of the cooling plant from July 1st to Oct. 15th, 2021 was simulated under three control strategies, as is shown in Table 3. Furthermore, the energy cost of the Mid-day24 control strategy was simulated to assess the extreme of control optimization models. The Mid-day24 strategy used the Mid-day24 prediction results and optimized the strategy at each hour of the day with the rule-based method. According to the simulation results, the operation cost under the Fixed schedule was 923,341 RMB. The Day-ahead prediction-based control costs were 840,685 RMB, which reduced 9.0 % of the cost (82,656 RMB) compared to the Fixed schedule. With the modification of Mid-day6 prediction-based control, the cost was reduced by 91,528 RMB (9.9 %) compared to the Fixed schedule. The electricity cost under the Mid-day6 control was 831,813 RMB, suggesting a cost benefit

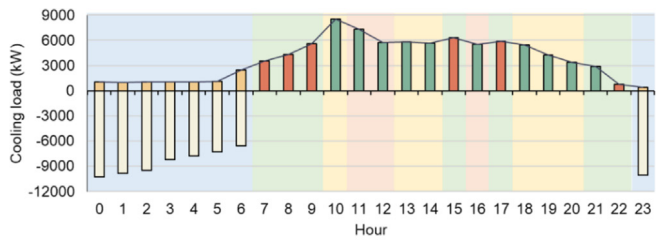

*(a) Fixed schedule on July 21st, 2021*

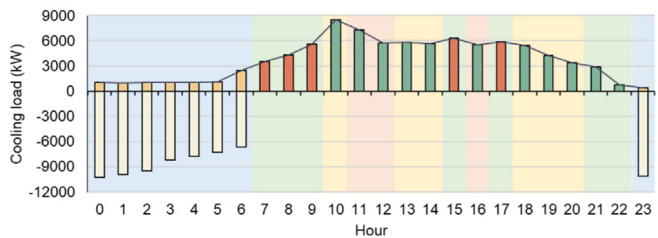

*(b) Day-ahead prediction-based control on July 21st, 2021*

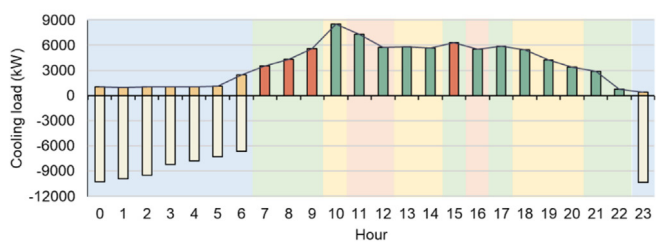

*(c) Mid-day6 prediction-based control on July 21st, 2021*

**Fig. 13.** Results of Fixed schedule (a), Day-ahead prediction-based control (b), and Mid-day6 prediction-based control (c) on July 21st, 2021.

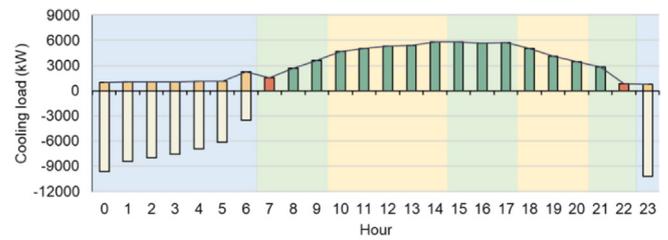

*(a) Fixed schedule on Sept. 7th, 2021*

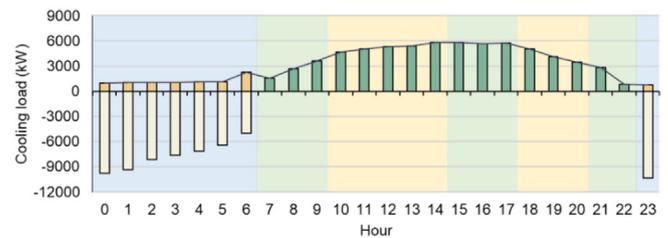

*(b) Day-ahead prediction-based control on Sept. 7th, 2021*

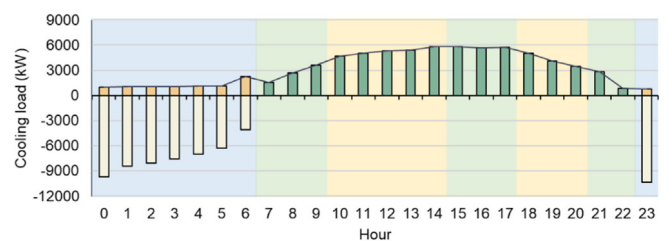

*(c) Mid-day6 prediction-based control on Sept. 7th, 2021*

**Fig. 14.** Results of Fixed schedule (a), Day-ahead prediction-based control (b), and Mid-day6 prediction-based control (c) on Sept. 7th, 2021.

from a mid-day modification. The energy cost saving rate of the Mid-day24 model was 10.0 %, which is only slightly higher than the Mid-day6 model. The trade-off usually occurs during partial-peak TOU hours. Mid-day modifications at early partial-peak hours were sufficient to reach a cost-effective optimal control solution. It is theoretically true that Mid-day24 model achieves slightly more cost savings than Mid-day6 model with acceptable consumptions of computation resources, and should be applied for fully automated control system. However, given that the control system of this case still requires manual confirmation and interference for every operation and adjustment, the Mid-day6 model achieved a high performance with relatively minor computation resources and manual efforts, and is considered the appropriate model for practical applications.

### 3.3. On-site deployment and the performance of the model

The proposed Mid-day6 prediction-based control method was formulated as a software and integrated into the BAS of the commercial complex. The software was installed on a local computer monitoring the operation of the cooling plant. The computer has access to the Internet to collect weather data from online APIs and communicates with the cooling plant by Modbus TCP/IP.

The software is composed of a cooling load prediction model and rule-based control method. The cooling load prediction model is trained in advance before each prediction window or prediction time point with the collected historical weather data and historical cooling load from cooling plant meters recursively. The well-trained cooling load prediction model acquires forecast weather data from ColorfulClouds [50] weather API for real-time cooling load predictions. The predicted cooling load is then fed to the rule-based control method. With predicted cooling load from the cooling load prediction model and ice storage state from the cooling plant measurement, the rule-based control method provides the suggested future hourly operation strategy to the TES system cooling plant. The chillers, pumps, valves, and cooling towers are operated according to the suggested operation strategy.

The software deployed in the commercial complex has been in operation since the start of the cooling season of 2022. The cooling load prediction performance of two weeks from June 27th to July 11st, 2022 is shown in Table 4. Mid-day6 model was adopted for the on-site operation. The on-site prediction CVMAE was 15.1 %, and the CVRMSE was 20.9 %. Compared to the errors of Mid-day6 model in Table 2, the proposed Mid-day6 model realizes similar accuracy for cooling load prediction in on-site operation. Simulation analysis of the cost reduction performance is conducted based on the on-site cooling load prediction result. As Table 5 shows, the Mid-day6 prediction-based control saves the operating cost of the cooling plant by 13,595 RMB in the two weeks compared to the Fixed schedule, with a cost reduction rate of 10.1 %. The proposed prediction-based control software brings expected on-site cost benefit for the commercial complex.

**Table 3**
Energy cost of Fixed schedule, Day-ahead control, Mid-day6 control and Mid-day24 control.

|  | Fixed schedule | Day-ahead control | Mid-day6 control | Mid-day24 control |
|---|---|---|---|---|
| Cost (RMB) | 923,341 | 840,685 | 831,813 | 830,548 |
| Cost reduction (RMB) | \ | 82,656 | 91,528 | 92,793 |
| Cost reduction rate | \ | 9.0 % | 9.9 % | 10.0 % |

**Table 4**
Cooling load prediction errors of the on-site operation from June 27th to July 11th, 2022.

|  | MAE (kW) | CVMAE | RMSE (kW) | CVRMSE |
|---|---|---|---|---|
| Prediction error | 506 | 15.1 % | 700 | 20.9 % |

**Table 5**
Control performance comparison of the on-site operation test from June 27th to July 11st, 2022.

|  | Cost (RMB) | Cost reduction (RMB) | Cost reduction rate |
|---|---|---|---|
| Fixed-schedule control (Original strategy) | 134,040 | \ | \ |
| Mid-day6 prediction-based control (Proposed strategy) | 120,445 | 13,595 | 10.1 % |

## 4. Discussion

### 4.1. Application-oriented evaluation mechanism for prediction models

The above-mentioned sections propose an approach of cooling load prediction and optimal control strategy for ice-based TES systems. Energy load prediction in buildings is a hot topic that has attracted extensive research interest in the past few decades. Researchers have devoted massive efforts to train and improve the predicting algorithms to improve prediction accuracy. The evaluation metrics, however, are generally consistent and identical. Most of the current prediction studies use statistical metrics such as MAE, RMSE, mean absolute percentage error, mean bias error (MBE), r-square, or coefficient of variance [39]. The optimization and selection of the model are solely based on these statistics and major efforts are put into minimizing the gap between hourly predictions and observations, which is sufficiently small and acceptable.

However, very limited studies have evaluated the prediction model from applicational perspectives. Few studies introduce engineering domain metrics to evaluate or optimize prediction models to fit for different engineering purposes. It is notable that reducing the hourly prediction errors may have little effect on final applicational results. Taking the results of the proposed model in this study as an example, Table 2 implicates that the predicting error of the Mid-day24 model (CVMAE = 10.9 %) is evidently lower than the Mid-day6 model (CVMAE = 12.5 %). However, the energy cost saving based on the optimal control of the two prediction models are essentially the same (Mid-day6: 9.9 %, Mid-day24: 10.0 %; Shown in Table 3). This is because the mid-day modifications at partial-peak hours (Mid-day6) have captured the key features in daily cooling load profiles and provides sufficient information for the decision making of the "trade-off" of ice storage usage at partial-peak hours. The prediction modifications at every hour of the day (Mid-day24) provide little additional information to support control optimizations. If the evaluation is solely based on prediction errors, the Mid-day24 model would be selected. However, from the applicational perspective of ice-based TES control, Mid-day6 model uses less operations and modifications to achieve essentially similar performance. From an engineering perspective, using less operations to achieve appreciable benefits can be reliable and robust.

The above-mentioned case serves as a reminder to evaluate prediction models from various perspectives, especially from applicational aspects. Apart from point-to-point predicting errors, the prediction model should be evaluated and optimized based on other metrics such as total loads during peak hours, energy costs, and thermal comfort dissatisfaction rate, etc. Previous research has laid solid foundations for accurate predicting algorithms for energy forecasts and analysis. Future efforts should focus on how prediction models connect to the specific applications and how they should be evaluated and optimized to fit specific applicational purposes.

### 4.2. Limitations and future perspectives

The proposed model introduced a mid-day modification mechanism and control-oriented architecture for training and optimization of prediction models. The algorithms used in this research are commonly used well-tuned machine learning methods and packages. In future studies, the prediction model could be further improved with more advanced machine learning algorithms and techniques. Modern ensemble models such as LightGBM and deep learning models such as CNN and DBN could extract specific features (e.g. time-relevant features) in cooling load profiles and may have massive potential to improve the prediction accuracy in such cases. According to different optimization purposes, specific techniques can be implemented to design, train, and tune the machine learning models and may yield more practical results. Also, the impact of the uncertainty of the inputs on the predictions is also critical. Wang et al. [34] discussed the uncertainty of the weather forecasts and its impact on cooling load prediction models. The uncertainty of measurements in cooling plants, including temperature, flow rate and power measurements, should also be carefully considered. It is essential to at least conduct balancing validation at different points of the cooling plant system before the cooling load data could be used in subsequent analysis. Advanced models that evaluate the uncertainty of the predicting accuracy caused by the uncertainty from input weather data and cooling system measurements should also be developed and analyzed.

For control optimization, the current control strategy only focused on the allocation of ice release during the day. The optimization of ice storage strategy during the night is also an impor-tant issue, especially considering heat loss of the ice storage tank. The whole-day control strategy of the ice-based TES should be further investigated. Moreover, the proposed research used a simplified rule-based method. The rule-based method relies largely on the TOU tariff characteristics, and may not be applicable under dynamic pricing schemes. Future research may introduce more advanced control optimization methods. Reinforcement learning can be a powerful solution as it can use TOU tariff as inputs and realize automated control optimization, especially under dynamic energy tariffs. Implementing reinforcement learning or other advanced control algorithms is a promising future research perspective for optimal TES system control.

## 5. Conclusion

TES is an efficient technical approach for demand response and load shifting in commercial buildings. Management and allocation of daily cooling loads is a major area in which to improve the performance of TES, which relies on cooling load predictions and optimized control strategy. Studies have conducted numerous comprehensive investigations on cooling load predictions and control models individually. However, the gap still remains in the integrated optimization of prediction and control model combined, and in practical applications of such models in TES operations. This research proposed a novel data-driven approach for cooling load prediction and prediction-based optimal control for ice-based TES in commercial buildings. A cooling load prediction model was developed to predicting algorithm selection, input parameter combinations, and mechanism optimizations. A mid-day modification was introduced into the prediction model to improve the prediction accuracy according to the TOU tariff profiles. Based on these predictions, a rule-based control strategy was proposed, and the mid-day adjustment was introduced in correspondence with mid-day prediction modifications to improve the control performance of the model. The proposed method was applied to an ice-based TES of a commercial complex in Beijing. During the cooling season of 2021, the predicting MAE of the proposed model was 389 kW, with CVMAE at 12.5 %. The RMSE of the prediction was 599 kW, and the CVRMSE was 19.2 %. The proposed prediction-based control model saves an energy cost of RMB 91,528, with a cost saving rate of 9.9 % compared to that of the current Fixed-schedule operations. The proposed model was implemented as a software and has been deployed into the BAS of the case building, achieving significant improvements for TES management. The key of this study is to propose an integrated approach that combines cooling load prediction and control strategy optimizations, improves the real-time performance with mid-day modification mechanism, and evaluates the applicational performance in practical cases. The enhancement of prediction models, control models, and the model integration can improve the degree of automation in TES and achieve a better overall efficiency of the cooling plants, supporting practical applications of energy-flexible HVAC systems for commercial buildings.

## Data availability

No data was used for the research described in the article.

## Declaration of Competing Interest

The authors declare that they have no known competing financial interests or personal relationships that could have appeared to influence the work reported in this paper.

## Acknowledgements

This publication has been jointly written within the cooperative project "Key technologies and demonstration of combined cooling, heating and power generation for low-carbon neighbourhoods/ buildings with clean energy – ChiNoZEN". The authors gratefully acknowledge the funding support from the Ministry of Science and Technology of China (MOST project number 2019YFE0104900), and from the Research Council of Norway (NRC project number 304191 - ENERGIX). The authors would also like to acknowledge the support from Beijing Municipal Natural Science Foundation of China (Grant number 8222019) and Tsinghua-Foshan Innovation Special Fund (TFISF 2021THFS0201).

## Appendix A. Prediction results of all 127 cases in input parameter sensitivity analysis.

**Table A1**
Prediction results of all input feature combinations for the building.

| Input feature combination | Mall MAE (kW) | Office MAE (kW) | Input feature combination | Mall MAE (kW) | Office MAE (kW) |
|---|---|---|---|---|---|
| T | 1061 | 617 | THRW | 870 | 267 |
| H | 1380 | 665 | THRO | 541 | 443 |
| R | 1257 | 563 | THRL | 351 | 420 |
| S | 1659 | 708 | THSW | 978 | 421 |
| W | 1647 | 602 | THSO | 540 | 444 |
| O | 972 | 520 | THSL | 369 | 429 |
| L | 379 | 446 | THWO | 557 | 216 |
| TH | 966 | 591 | THWL | 366 | 274 |
| TR | 893 | 490 | THOL | 348 | 414 |
| TS | 1045 | 619 | TRSW | 903 | 289 |
| TW | 1079 | 457 | TRSO | 556 | 445 |
| TO | 584 | 435 | TRSL | 357 | 415 |
| TL | 367 | 438 | TRWO | 565 | 224 |
| HR | 1115 | 521 | TRWL | 356 | 219 |
| HS | 1394 | 659 | TROL | 352 | 406 |
| HW | 1363 | 533 | TSWO | 598 | 225 |
| HO | 843 | 503 | TSWL | 376 | 270 |
| HL | 381 | 445 | TSOL | 352 | 407 |
| RS | 1247 | 549 | TWOL | 355 | 210 |
| RW | 1247 | 377 | HRSW | 1124 | 331 |
| RO | 865 | 502 | HRSO | 777 | 485 |

**Table A1** (continued)

| Input feature combination | Mall MAE (kW) | Office MAE (kW) | Input feature combination | Mall MAE (kW) | Office MAE (kW) |
|---|---|---|---|---|---|
| RL | 371 | 414 | HRSL | 373 | 418 |
| SW | 1646 | 595 | HRWO | 749 | 278 |
| SO | 965 | 517 | HRWL | 373 | 236 |
| SL | 381 | 446 | HROL | 366 | 411 |
| WO | 904 | 324 | HSWO | 794 | 298 |
| WL | 386 | 293 | HSWL | 386 | 287 |
| OL | 372 | 413 | HSOL | 372 | 415 |
| THR | 860 | 486 | HWOL | 378 | 236 |
| THS | 964 | 585 | RSWO | 864 | 318 |
| THW | 978 | 429 | RSWL | 379 | 238 |
| THO | 539 | 446 | RSOL | 371 | 408 |
| THL | 362 | 438 | RWOL | 375 | 224 |
| TRS | 896 | 492 | SWOL | 379 | 235 |
| TRW | 901 | 292 | THRSW | 871 | 269 |
| TRO | 544 | 439 | THRSO | 541 | 444 |
| TRL | 356 | 419 | THRSL | 354 | 419 |
| TSW | 1060 | 448 | THRWO | 559 | 217 |
| TSO | 590 | 441 | THRWL | 355 | 219 |
| TSL | 369 | 432 | THROL | 348 | 412 |
| TWO | 594 | 220 | THSWO | 557 | 216 |
| TWL | 373 | 268 | THSWL | 369 | 275 |
| TOL | 354 | 407 | THSOL | 354 | 409 |
| HRS | 1131 | 519 | THWOL | 354 | 215 |
| HRW | 1114 | 327 | TRSWO | 572 | 225 |
| HRO | 756 | 480 | TRSWL | 358 | 216 |
| HRL | 371 | 418 | TRSOL | 353 | 406 |
| HSW | 1372 | 534 | TRWOL | 352 | 204 |
| HSO | 853 | 503 | TSWOL | 359 | 213 |
| HSL | 379 | 437 | HRSWO | 755 | 282 |
| HWO | 796 | 297 | HRSWL | 377 | 237 |
| HWL | 384 | 290 | HRSOL | 368 | 409 |
| HOL | 376 | 417 | HRWOL | 374 | 223 |
| RSW | 1237 | 376 | HSWOL | 379 | 238 |
| RSO | 876 | 499 | RSWOL | 379 | 226 |
| RSL | 373 | 415 | THRSWO | 557 | 218 |
| RWO | 866 | 316 | THRSWL | 353 | 218 |
| RWL | 375 | 239 | **THRSOL** | **343** | 411 |
| ROL | 368 | 407 | **THRWOL** | 351 | **203** |
| SWO | 891 | 326 | THSWOL | 356 | 213 |
| SWL | 387 | 287 | TRSWOL | 351 | 205 |
| SOL | 371 | 412 | HRSWOL | 375 | 225 |
| WOL | 377 | 232 | THRSWOL | 352 | 207 |
| THRS | 860 | 484 | | | |